# Determination of the magnetic field dependence of the surface resistance of superconductors from cavity tests

J. R. Delayen,[1,2,*] H. Park,[1,2] S. U. De Silva,[1] G. Ciovati,[1,2] and Z. Li[3]

[1]*Center for Accelerator Science, Department of Physics, Old Dominion University, Norfolk, Virginia 23529, USA*
[2]*Thomas Jefferson National Accelerator Facility, Newport News, Virginia 23606, USA*
[3]*SLAC National Accelerator Laboratory, Menlo Park, California 94025, USA*

We present a general method to derive the magnetic field dependence of the surface resistance of superconductors from the $Q$-curves obtained during the cryogenic tests of cavities. The results are applied to coaxial half-wave cavities, TM-like "elliptical" accelerating cavities, and cavities of more complicated geometries.

## I. INTRODUCTION

The well-known expressions for the surface resistance of superconductors in electromagnetic fields, and its dependence on frequency, temperature, and a few materials parameters, were obtained as a perturbation theory under the assumption that the magnitude of the electromagnetic field is much smaller than the critical field [1–4]. This resulted in a surface resistance independent of the magnitude of the electromagnetic field. There have been several attempts at developing theories of the surface resistance at high rf field [5–7], up to the critical field, but, at present, there is no universally accepted consensus on the correct fully self-consistent theory.

Experimentally, cryogenic tests of superconducting cavities developed for particle accelerators have shown that superconductors can display a strong dependence of their surface resistance on the rf field. Furthermore, the field dependence can vary greatly depending on the history of the cavities: chemical treatment, high temperature and low temperature [8,9] heat treatment, impurity concentration [10,11], ambient magnetic field during transition [12], cooling rate during transition [13], and so on. Traditionally superconductors have shown an increase of their surface resistance with rf field [14,15] which puts a limit to the accelerating gradients achievable for high-energy accelerators. More recently, processes have been developed that yield a decrease of the surface resistance with medium fields [16,17], a great benefit for accelerators operating in cw mode. A strong dependence of the surface resistance on magnetic field is also often observed in cavities made by sputtering of Nb on Cu [18,19] or in cavities made of $Nb_3Sn$ on Nb [20]. Developing a full understanding and theory of the rf field dependence of the surface resistance of superconductors will require accurate knowledge of that surface resistance as a function of the rf field, and its dependence on preparation and processing parameters. Experimentally it has proven difficult to develop techniques where a superconductor was exposed to a uniform electromagnetic field. Until now, in all measurements, either the superconducting sample was exposed to a non-uniform field in a test cavity or, more often, an entire cavity was made of superconducting material that was exposed to a field that ranged from 0 to a maximum value.

In this paper we present a method and formulae that allow determination of the actual dependence of the surface resistance from experiments where an "average" surface resistance is derived from tests of superconducting resonators. An underlying assumption is still that, while the surface resistance has a magnetic field dependence, it does not have a dependence on location. It is not always true as it is known that superconducting cavities can have "hot spots" where the surface resistance is higher and with a stronger field dependence than in the rest of the cavity [21,22].

## II. ANALYTICAL METHOD

A cryogenic test of a superconducting cavity often consists of the measurement of the quality factor $Q$ as a function of some field, either the average accelerating field or a peak surface field. Here we assume that we experimentally measure $Q(H_p)$ as a function of the peak surface magnetic field $H_p$.

*jdelayen@odu.edu

A characteristic property of an electromagnetic mode of a cavity is its geometrical factor $G$ defined as

$$G = \omega \mu_0 \frac{\int_V |\mathbf{H}|^2 dV}{\int_S |\mathbf{H}|^2 dS}.$$ (1)

It depends only on the shape of the cavity and the electromagnetic mode; it is independent of material and frequency (or size).

If the surface resistance $R_s$ is constant, then $G = QR_s$. While the assumption of constant surface resistance may be valid for normal conductors it often is not for superconductors as their surface resistance can display a strong dependence on the local surface magnetic field.

Dividing the geometrical factor $G$ by the measured $Q(H_p)$ we obtain a surface resistance $\bar{R}_s(H_p)$. Since the magnetic field is not constant over the whole surface, $\bar{R}_s$ is only an average surface resistance. The goal is to obtain the actual $R_s(H)$ from the experimentally measured $\bar{R}_s(H)$. The two are related by

$$\bar{R}_s(H_p) \int_S |\mathbf{H}(\vec{r})|^2 dS = \int_S R_s(H(\vec{r})) |\mathbf{H}(\vec{r})|^2 dS,$$ (2)

where the integrals are taken over the whole cavity area.

We define the function $a(h)$ as the fraction of the total cavity area where $|H| \leq hH_p$ when $H_p$ is the peak surface magnetic field in the cavity. Clearly $a(h)$ is a continuous monotonically increasing function with $a(0) = 0$ and $a(1) = 1$. Furthermore, for many of the normally shaped cavities with some degree of symmetry, the peak surface magnetic field $H_p$ does not occur at a single point but on a closed contour. This is the case for example for quarter-wave, half-wave, spoke, and $TM_{010}$ type cavities. In that case we have $\frac{da}{dh}|_{h=1} = \infty$. For more complex geometries where $H_p$ occurs at a single point $\frac{da}{dh}|_{h=1}$ can remain finite but still take large values close to $h = 1$.

Alternatively, $a(h)$ can be interpreted as the probability distribution for the surface magnetic field and $\frac{da}{dh}$ as its probability density.

Because of the continuity and monotonicity of $a(h)$ we can make a change of variable in the integrals in Eq. (2) and integrate over the magnetic field instead of over the area.

$$\bar{R}_s(H) \int_0^1 (hH)^2 \frac{da}{dh} dh = \int_0^1 R_s(hH)(hH)^2 \frac{da}{dh} dh.$$ (3)

We now assume that the experimentally measured $\bar{R}_s(H)$ can be expanded in a sum of powers of the magnetic field

$$\bar{R}_s\left(\frac{H}{H_0}\right) = \bar{R}_0 \sum_{\alpha_i} r_{\alpha_i} \left(\frac{H}{H_0}\right)^{\alpha_i}.$$ (4)

The sum, which can be of any length, is not restricted to integer powers as in Taylor series. The coefficients $\alpha_i$ can be any non-negative real numbers and can be chosen to provide a best approximation to the experimental data. For the sake of convenience we assume that $\alpha_0 = 0$ and that the suite is ordered ($\alpha_i < \alpha_j$ if $i < j$).

The magnetic field $H_0$ is arbitrary and is introduced to make the coefficients $r_{\alpha_i}$ dimensionless. $\bar{R}_0$ is the zero-field surface resistance and, since $\alpha_0 = 0$, $r_{\alpha_0} = 1$.

We assume the same power expansion for the actual surface resistance but with the coefficients modified by the factors $\beta(\alpha_i)$

$$R_s\left(\frac{H}{H_0}\right) = R_0 \sum_{\alpha_i} \beta(\alpha_i) r_{\alpha_i} \left(\frac{H}{H_0}\right)^{\alpha_i},$$ (5)

with $R_0 = \bar{R}_0$ and $\beta(\alpha_0) = 1$.

Replacing $\bar{R}_s(H/H_0)$ and $R_s(H/H_0)$ by their expansions in the integrals in Eq. (3) and equating identical powers of $(H/H_0)$ we obtain the factors $\beta(\alpha_i)$ relating the coefficients in the power expansion of the average and actual surface resistance. The functional dependence of the correction coefficients $\beta(\alpha)$ is given by

$$\beta(\alpha) = \frac{\int_0^1 h^2 \frac{da}{dh} dh}{\int_0^1 h^{2+\alpha} \frac{da}{dh} dh}.$$ (6)

Since $a(h)$ is monotonically increasing $\frac{da}{dh} \geq 0$, and since $0 \leq h \leq 1$, $\beta(\alpha_i) > \beta(\alpha_j)$ for $\alpha_i > \alpha_j$. The function $\beta(\alpha)$ is a smooth, continuous, monotonically increasing function with $\beta(0) = 1$, so it needs to be calculated only for a small number of $\alpha$ and its value can be obtained for any other by interpolation.

A hypothetical cavity with a uniform surface magnetic field would have $a(0 \leq h < 1) = 0$, $a(1) = 1$, and $\frac{da}{dh} = \delta(1-h)$, which would give from Eq. (6) $\beta(\alpha) = 1$, $\forall \alpha$ as expected.

For more complicated cavity geometries where $a(h)$ cannot be obtained analytically it may be more convenient to use an equivalent relationship obtained by performing an integration by parts in the above expression for $\beta(\alpha)$.

$$\int_0^1 h^{2+\alpha} \frac{da}{dh} dh = h^{2+\alpha} a(h)\Big|_0^1 - (2+\alpha) \int_0^1 h^{1+\alpha} a(h) dh$$

$$= 1 - (2+\alpha) \int_0^1 h^{1+\alpha} a(h) dh$$

$$= (2+\alpha) \int_0^1 h^{1+\alpha} [1 - a(h)] dh,$$ (7)

which yields

$$\beta(\alpha) = \frac{2 \int_0^1 h[1 - a(h)] dh}{(2+\alpha) \int_0^1 h^{1+\alpha} [1 - a(h)] dh}.$$ (8)

In the two integrals in the above equation the quantity in brackets now represents the fraction of the cavity area that sustains a magnetic field $|H| > hH_p$.

Conceptually it would be even possible to assume a continuum spectrum of exponents $\alpha$ where the surface resistances would be of the form

$$\bar{R}_s\left(\frac{H}{H_0}\right) = \bar{R}_0 \int_0^\infty r(\alpha)\left(\frac{H}{H_0}\right)^\alpha d\alpha,$$

$$R_s\left(\frac{H}{H_0}\right) = R_0 \int_0^\infty \beta(\alpha) r(\alpha)\left(\frac{H}{H_0}\right)^\alpha d\alpha. \quad (9)$$

$r(\alpha)$ would then be related to $\bar{R}_s$ through a Laplace transform and its inverse, and the factors $\beta(\alpha)$ would still be given by Eqs. (6) or (8). It is not clear that such a complication would be beneficial in practical applications for realistic cavities and superconductors.

## III. APPLICATION TO COAXIAL HALF-WAVE CAVITY

The above results are now applied to a coaxial half-wave cavity of length $L$, center conductor radius $a$, and outer conductor radius $b$, operating in any of the TEM modes. We define the dimensionless parameters $\rho = a/b$ and $\delta = b/L$.

Such a cavity, shown in Fig. 1, has been built and is being used specifically for the investigation of the frequency, rf field, and temperature dependence of the surface resistance of superconductors [23,24]. Its dimensions are $a = 19.5$ mm, $b = 101$ mm, and $L = 460$ mm.

The area of the center conductor is $A_1 = 2\pi a L = 2\pi\rho\delta L^2$. The peak magnetic field on the center conductor is $H_p$ and the area that sustains a magnetic field $|H| \le hH_p$ is

$$A_1(h) = 4\rho\delta L^2 \arcsin(h), \quad (10)$$

$$\frac{dA_1(h)}{dh} = 4\rho\delta\frac{L^2}{\sqrt{1-h^2}}. \quad (11)$$

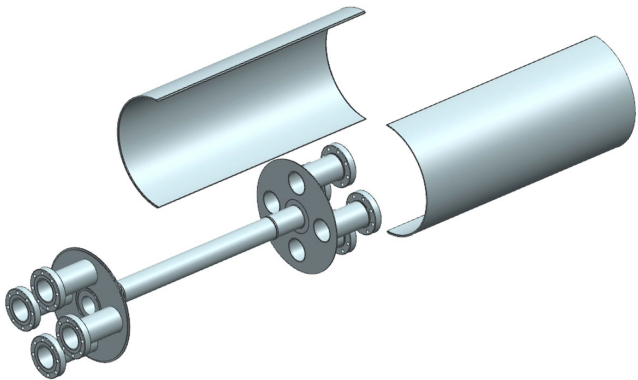

FIG. 1.   Exploded view of the coaxial half-wave cavity.

The area of the outer conductor is $A_2 = 2\pi b L = 2\pi\delta L^2$. The peak magnetic field on the outer conductor is $\rho H_p$ and the area that sustains a magnetic field $|H| \le hH_p$ is

$$A_2(h) = \begin{cases} 4\delta L^2 \arcsin(h/\rho) & \text{if } 0 \le h < \rho, \\ 2\pi\delta L^2 & \text{if } \rho \le h \le 1, \end{cases} \quad (12)$$

$$\frac{dA_2(h)}{dh} = \begin{cases} \dfrac{4\delta}{\rho}\dfrac{L^2}{\sqrt{1-(h/\rho)^2}} & \text{if } 0 \le h < \rho, \\ 0 & \text{if } \rho \le h \le 1. \end{cases} \quad (13)$$

The area of the two end plates is $2\pi(b^2 - a^2) = 2\pi\delta^2 L^2(1 - \rho^2)$. They sustain a maximum magnetic field $H_p$ and a minimum magnetic field $\rho H_p$. The area where the magnetic field is $|H| \le hH_p$ is

$$A_3(h) = \begin{cases} 0 & \text{if } 0 \le h < \rho, \\ 2\pi\delta^2 L^2(1 - (\rho/h)^2) & \text{if } \rho \le h \le 1, \end{cases} \quad (14)$$

$$\frac{dA_3(h)}{dh} = \begin{cases} 0 & \text{if } 0 \le h < \rho, \\ \dfrac{4\pi\rho^2\delta^2 L^2}{h^3} & \text{if } \rho \le h \le 1. \end{cases} \quad (15)$$

The contributions of the inner conductor, outer conductor, and end plates to the function $a(h)$ and the function $a(h)$ itself are shown for a coaxial half-wave cavity with $b/L = 0.25$ in Fig. 2 for $a/b = 0.2$, and in Fig. 3 for $a/b = 0.5$. The functions $\frac{da}{dh}$ for the same cavities are shown in Fig. 4 for $a/b = 0.2$, and in Fig. 5 for $a/b = 0.5$ In both cases the singularities at $h = \rho$ and $h = 1$ are clearly visible.

From this we can calculate the coefficients in the expansion of the surface resistance in powers of the magnetic field using Eq. (6):

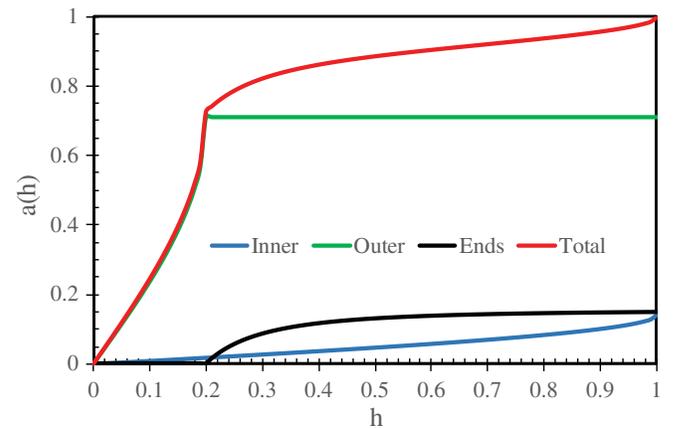

FIG. 2.   Fractional areas of a coaxial half-wave cavity where $H \le hH_p$, and contributions from the cavity components. The ratio of outer radius to length is 0.25 and the ratio of inner to outer radii is 0.2.

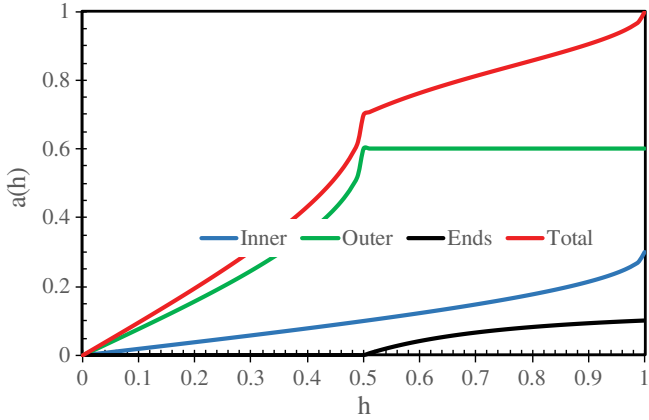

FIG. 3.  Fractional areas of a coaxial half-wave cavity where $H \leq hH_p$, and contributions from the cavity components. The ratio of outer radius to length is 0.25 and the ratio of inner to outer radii is 0.5.

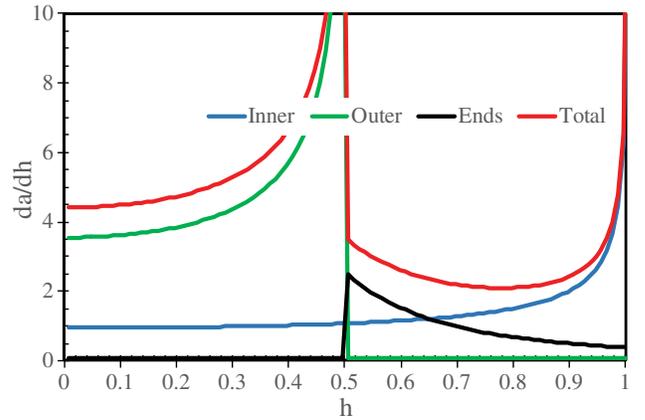

FIG. 5.  Derivative of the fractional areas of a coaxial half-wave cavity where $H \leq hH_p$, and contributions from the cavity components. The ratio of outer radius to length is 0.25 and the ratio of inner to outer radii is 0.5.

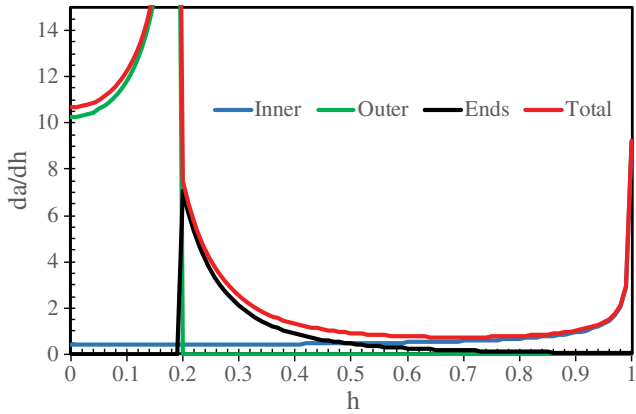

FIG. 4.  Derivative of the fractional areas of a coaxial half-wave cavity where $H \leq hH_p$, and contributions from the cavity components. The ratio of outer radius to length is 0.25 and the ratio of inner to outer radii is 0.2.

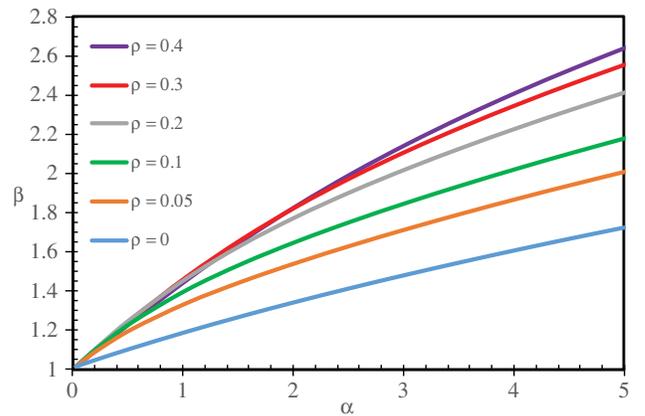

FIG. 6.  Dependence of the correction coefficients $\beta$ on the exponents $\alpha$ for various ratios of inner to outer conductor radii of a coaxial half-wave cavity with $b/L = 0.25$.

$$\beta(\alpha) = \frac{\rho \int_0^1 \frac{h^2}{\sqrt{1-h^2}} dh + \rho^{-1} \int_0^\rho \frac{h^2}{\sqrt{1-(h/\rho)^2}} dh + \int_\rho^1 \pi \rho^2 \delta h^{-1} dh}{\rho \int_0^1 \frac{h^{2+\alpha}}{\sqrt{1-h^2}} dh + \rho^{-1} \int_0^\rho \frac{h^{2+\alpha}}{\sqrt{1-(h/\rho)^2}} dh + \int_\rho^1 \pi \rho^2 \delta h^{-1+\alpha} dh} = \frac{(1+\rho)/4 + \rho\delta \ln(1/\rho)}{\frac{1+\rho^{1+\alpha}}{2\sqrt{\pi}} \frac{\Gamma(\alpha/2 + 3/2)}{\Gamma(\alpha/2 + 2)} + \frac{\rho\delta}{\alpha}(1-\rho^\alpha)}. \quad (16)$$

The factors $\beta$ calculated from Eq. (16) are shown as function of $\alpha$ for various $\rho$ in Fig. 6, and as function of $\rho$ for various $\alpha$ in Fig. 7. The latter clearly shows that the coefficients $\beta(\alpha)$ first increase as a function of $\rho$ and then decrease. Actually, for all $\alpha$, $\beta(\alpha)$ takes the same value for $\rho = 0$ and $\rho = 1$.

The curve corresponding to $\rho = 0$ in Fig. 6 is equivalent to assuming that the behavior is dominated by the center conductor and that only that area needs to be taken into consideration. In that case the coefficients $\beta(\alpha)$ are given by

$$\beta(\alpha) \simeq \frac{\sqrt{\pi}}{2} \frac{\Gamma(\alpha/2 + 2)}{\Gamma(\alpha/2 + 3/2)}, \quad (17)$$

However, it is clear from Fig. 6 that, even for very small $\rho$, $\beta(\alpha)$ is quite different from that given for $\rho = 0$. This means that the whole cavity area needs to be taken into consideration and the outer conductor and end plates need to be included.

These results are now applied to the half-wave cavity shown in Fig. 1. Figure 8 shows a typical curve of the quality factor $Q$ as function of the peak surface magnetic field obtained at 4.35 K when the cavity is operated in the 325 MHz fundamental mode. In that mode, the geometrical factor is $G = 57\,\Omega$, from which the average surface resistance $\bar{R}_s$ is obtained and shown as the dots in Fig. 9.

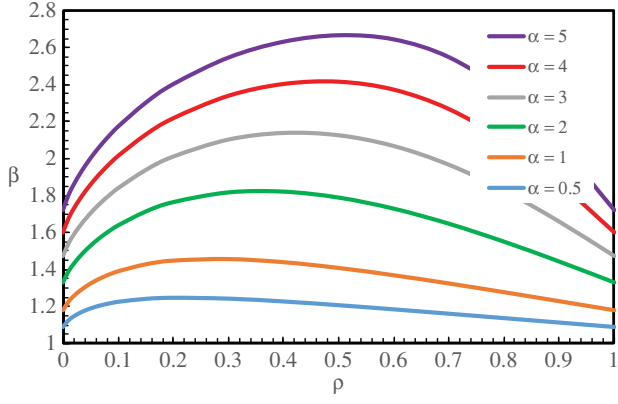

FIG. 7. Dependence of the correction coefficients $\beta$ on the ratios of inner to outer conductor radii of a coaxial half-wave cavity with $b/L = 0.25$ for various exponents $\alpha$.

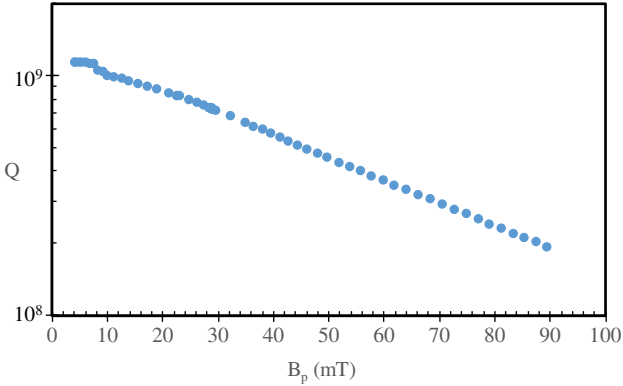

FIG. 8. $Q$-curve as function of peak surface magnetic field for the cavity shown in Fig. 1 operated in the fundamental 325 MHz mode at 4.32 K.

A polynomial fit to the experimental data, shown as the green line going through the dots in Fig. 9 is

$$\bar{R}_s\left(\frac{B}{B_0}\right) = 48.2\left[1 + 1.54\left(\frac{B}{B_0}\right) \right.$$
$$\left. + 2.03\left(\frac{B}{B_0}\right)^2 + 3.20\left(\frac{B}{B_0}\right)^3\right], \quad (18)$$

where we have chosen $B_0 = 100$ mT and the surface resistance is expressed in n$\Omega$.

From Eq. (16) the correcting factors $\beta(\alpha)$ are

$$\beta(0) = 1, \quad \beta(1) = 1.45, \quad \beta(2) = 1.76, \quad \beta(3) = 2.01,$$
$$(19)$$

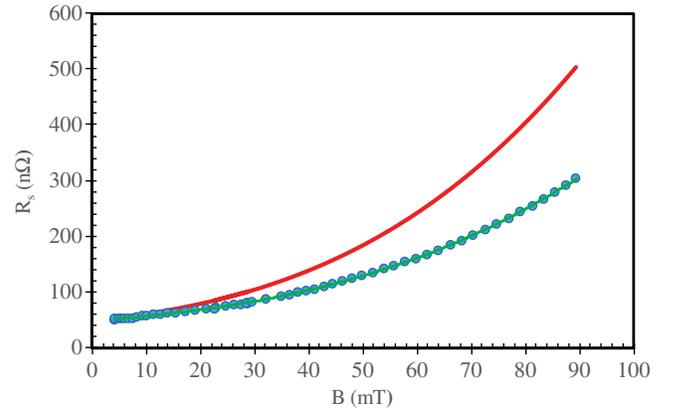

FIG. 9. Experimental average surface resistance obtained from $Q$-curve in Fig. 8 (blue dots) with polynomial fit given by Eq. (18) (green line) and the derived real surface resistance given by Eq. (20) (red line).

from which we obtain the actual surface resistance

$$R_s\left(\frac{B}{B_0}\right) = 48.2\left[1 + 2.23\left(\frac{B}{B_0}\right) \right.$$
$$\left. + 3.58\left(\frac{B}{B_0}\right)^2 + 6.42\left(\frac{B}{B_0}\right)^3\right]. \quad (20)$$

which is also shown in Fig. 9.

## IV. APPLICATION TO CAVITIES WITH AXIAL SYMMETRY

### A. TM$_{010}$ mode of a spherical cavity

Another example where the function $a(h)$ can be obtained analytically is a spherical cavity operating in the TM$_{010}$ mode where the magnetic field on the surface has a simple angular distribution: $H = H_p \sin\theta$ where $\theta$ is the angle with respect to the axis. The fractional area $a(h)$ where $|H| \leq H_p$ and its derivative are easily obtained:

$$a(h) = 1 - \sqrt{1 - h^2}$$
$$\frac{da}{dh} = \frac{h}{\sqrt{1 - h^2}} \quad (21)$$

The coefficients $\beta(\alpha)$ relating the power expansion of the measured and real surface resistance are then

$$\beta(\alpha) = \frac{\int_0^1 \frac{h^3}{\sqrt{1-h^2}}\,dh}{\int_0^1 \frac{h^{3+\alpha}}{\sqrt{1-h^2}}\,dh} = \frac{4}{3\sqrt{\pi}}\frac{\Gamma(\alpha/2 + 5/2)}{\Gamma(\alpha/2 + 2)}, \quad (22)$$

and shown in Fig. 10.

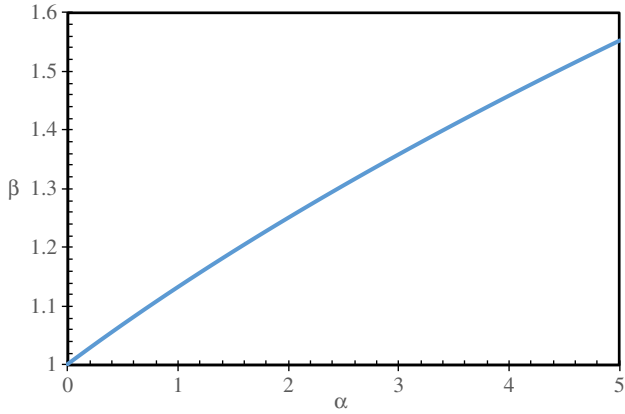

FIG. 10. Dependence of the correction coefficients $\beta$ on the exponents $\alpha$ for a spherical cavity operating in the $TM_{010}$ mode.

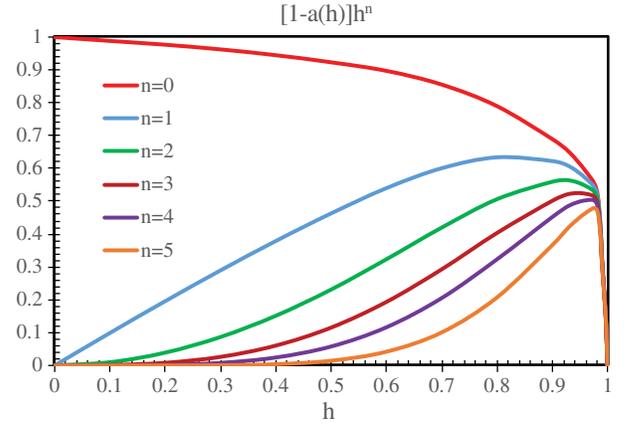

FIG. 12. Fraction of the surface area of the CEBAF original cavity shape where the surface magnetic field is larger than the fraction $h$ of the peak field and its product with several powers of $h$.

## B. $TM_{010}$ accelerating cavities and cavities with axial symmetry

For most cavities the function $a(h)$ cannot be obtained analytically but numerically. This is still relatively easy for cavities whose geometry displays axial symmetry. This is the case for the so-called "elliptical" $TM_{010}$ accelerating cavities and some coaxial quarter-wave and half-wave cavities.

An example is the original CEBAF cavity. The shape of the cavity and its surface magnetic field in the $TM_{010}$ mode obtained from Superfish [25] are shown in Fig. 11.

The function $a(h)$ can then be obtained by a simple numerical line integration of the shape and surface field profiles. Because $a(h)$ is obtained numerically, it is more practical to use Eq. (8) to calculate the correcting coefficients $\beta(\alpha)$. The function $[1 - a(h)]$ and its product with several powers of $h$ is shown in Fig. 12. The correcting coefficients $\beta(\alpha)$ are then calculated using Eq. (8). These are shown in Fig. 13 for the original CEBAF cavity.

As is clear from Fig. 11, that cavity was designed to have as constant magnetic field on its surface as possible in order to minimize the peak surface magnetic field. From Fig. 12 we see that the surface field is larger than 95% of its peak value over 50% of its area. We should expect, as is confirmed by Fig. 13, that the correcting coefficients $\beta(\alpha)$ should remain close to 1.

A $Q$-curve measured on that cavity [16] is shown in Fig. 14 and the experimentally-derived $\bar{R}_s(H)$ as the dots in Fig. 15.

A polynomial fit to $\bar{R}_s(H)$ is shown as the green line and is of the form

$$\bar{R}_s\left(\frac{B}{B_0}\right) = 10.3\left[1 - 1.61\left(\frac{B}{B_0}\right)\right.$$
$$\left. + 1.78\left(\frac{B}{B_0}\right)^2 - 0.58\left(\frac{B}{B_0}\right)^3\right], \quad (23)$$

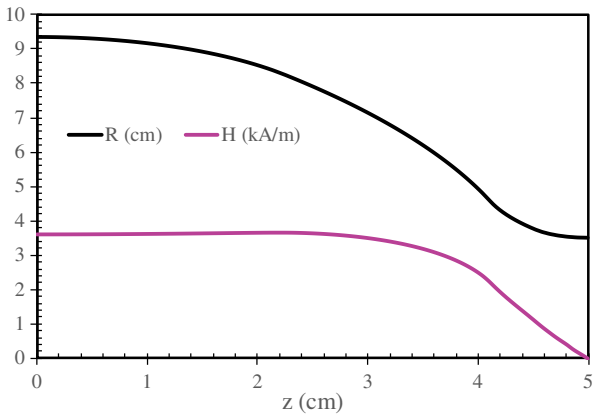

FIG. 11. Radius and surface magnetic field as function of distance on the axis from the midplane for the CEBAF original cavity shape.

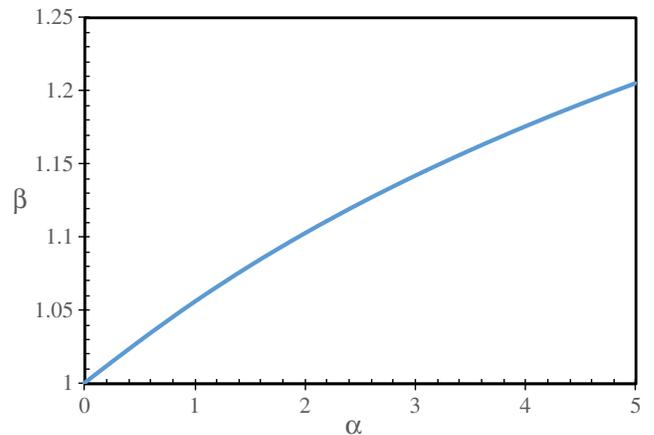

FIG. 13. Dependence of the correction coefficients $\beta$ on the exponents $\alpha$ for the CEBAF original cavity shape.

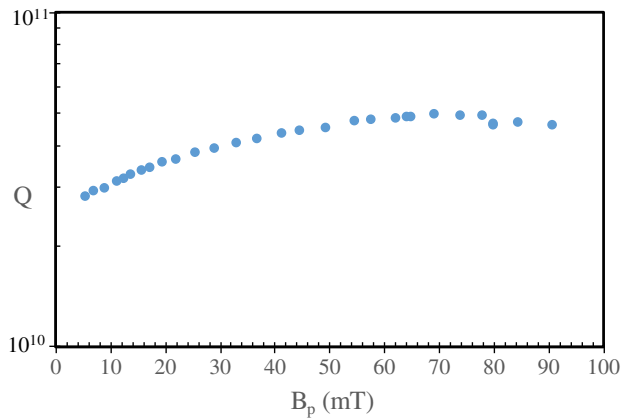

FIG. 14.    $Q$-curve of a CEBAF original shape cavity [16].

where we have chosen $B_0 = 100$ mT and the surface resistance is expressed in nΩ.

From Eq. (8) and Fig. 12 the correcting factors $\beta(\alpha)$ are

$$\beta(0) = 1, \quad \beta(1) = 1.06, \quad \beta(2) = 1.10, \quad \beta(3) = 1.14,$$ (24)

from which we obtain the actual surface resistance

$$R_s\left(\frac{B}{B_0}\right) = 10.3\left[1 - 1.71\left(\frac{B}{B_0}\right) \right. \\ \left. + 1.96\left(\frac{B}{B_0}\right)^2 - 0.66\left(\frac{B}{B_0}\right)^3\right].$$ (25)

which is also shown in Fig. 15.

As expected from the fact that the magnetic field is almost constant near its maximum value over a large fraction of the surface area, there is very little difference between $\bar{R}_s(H)$ and $R_s(H)$. Any difference would become apparent only for very strong dependence of the surface resistance on magnetic field.

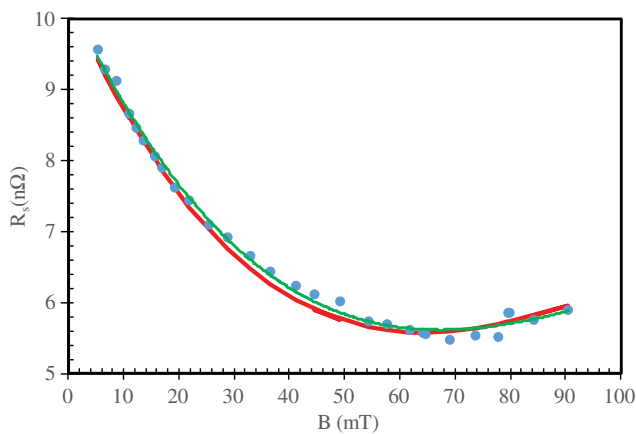

FIG. 15.    Experimental average surface resistance obtained from $Q$-curve in Fig. 14 (blue dots) with polynomial fit given by Eq. (23) (green line) and the derived real surface resistance given by Eq. (25) (red line).

## V. CAVITIES OF MORE COMPLEX GEOMETRY

In the case of cavities of more complicated geometry the distribution function of the surface magnetic field $a(h)$ must be obtained by sampling the magnetic field over the whole surface. This can be accomplished using the finite element field solver Omega3P [26]. Omega3P utilizes second-order curved tetrahedron elements and higher-order (up to order 6) field interpolation functions so that high accuracy can be achieved in rf field and geometry surface area calculations for the function $a(h)$.

Using the Omega3P solver, the minimum (0) and maximum ($H_p$) values of the surface magnetic field are obtained using the postprocessing tool. The fields are then normalized to $h_{max} = 1$. To calculate the function $a(h)$ the range $h \in [0, 1]$ is divided into a number $N$ of intervals. The $h$-field at each surface element is determined and added to the corresponding $h$-field bin. Because of the unstructured grid, the surface areas associated with the points on the surface are different and are determined by the element size they belong to. The surface areas of the points that fall into a $h$-bin are summed up to obtain the total surface $dS(h_i)$ associated with $h_i$. The function $a(h)$ is then obtained by a summation and normalization.

$$a(h_n) = \frac{\sum_1^n dS(h_i)}{\sum_1^N dS(h_i)}$$ (26)

By its very nature, the function $a(h)$ is defined only in the interval $h \in [0, 1]$. As mentioned previously, for normally shaped cavities, $\frac{da}{dh}|_{h=1}$ is either infinite or exhibits a large peak. Since the correction coefficients $\beta(\alpha)$ are mostly affected by the behavior of $a(h)$ near $h = 1$, especially for larger values of $\alpha$, the calculations need to include a large enough number of $h$-bins to resolve the singularity near $h = 1$. Similar singularities in $da/dh$ can also occur for other values of $h$—see for example $a(h)$ for the coaxial half-wave cavities in Figs. 2 and 3—but those are less important than the one at $h = 1$ since they contribute less to the integrals for the calculation of $\beta(\alpha)$.

An example of a cavity of more complicated geometry with no simple symmetry is a 400 MHz rf-dipole prototype cavity developed for the LHC High Luminosity Upgrade [27,28] shown in Fig. 16. For this particular cavity it was found that dividing $h \in [0, 1]$ into $N = 1600$ intervals was sufficient to provide the required accuracy. The "probability density" $da/dh$ for that cavity is shown in Fig. 17 and the distribution $a(h)$ in Fig. 18.

Since $a(h)$ is obtained numerically, the correction coefficients are more easily obtained using Eq. (8) which requires the integration of the products of $[1 - a(h)]$ with powers of $h$. These functions are shown in Fig. 19 for the rf-dipole cavity. The calculated correction coefficients $\beta(\alpha)$ are shown in Fig. 20.

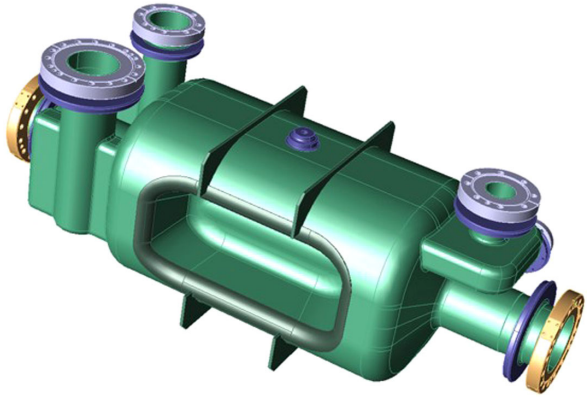

FIG. 16.   Design of a 400 MHz rf-dipole deflecting/crabbing cavity [27,28].

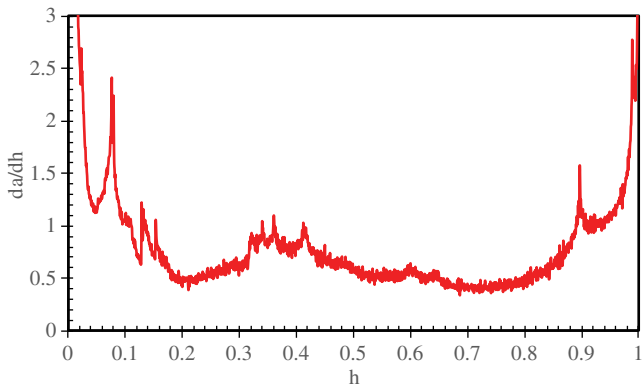

FIG. 17.   Function $da(h)/dh$ for the rf-dipole cavity shown in Fig. 16.

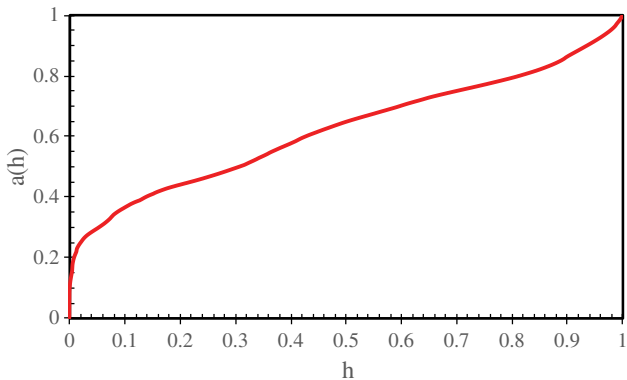

FIG. 18.   Fractional area $a(h)$ where $|H| \le hH_p$ for the rf-dipole cavity shown in Fig. 16.

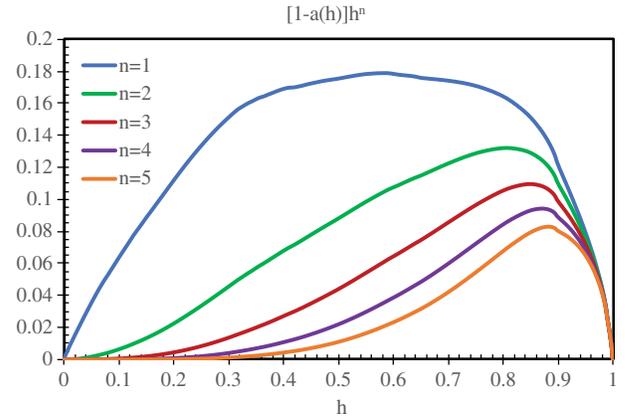

FIG. 19.   Products of the fraction of the surface area of the rf-dipole cavity, shown in Fig. 16, where the surface magnetic field is larger than the fraction $h$ of the peak field with several powers of $h$.

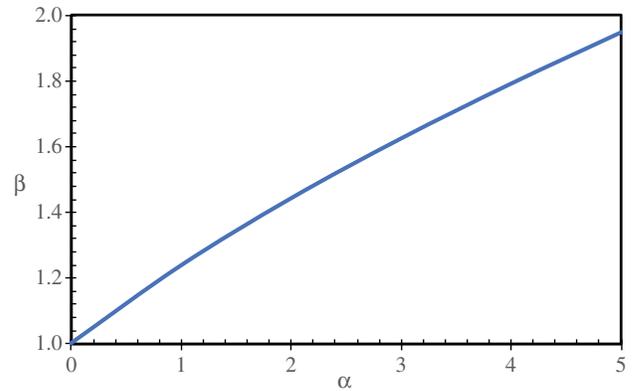

FIG. 20.   Dependence of the correction coefficients $\beta$ on the exponents $\alpha$ for the rf-dipole cavity shown in Fig. 16.

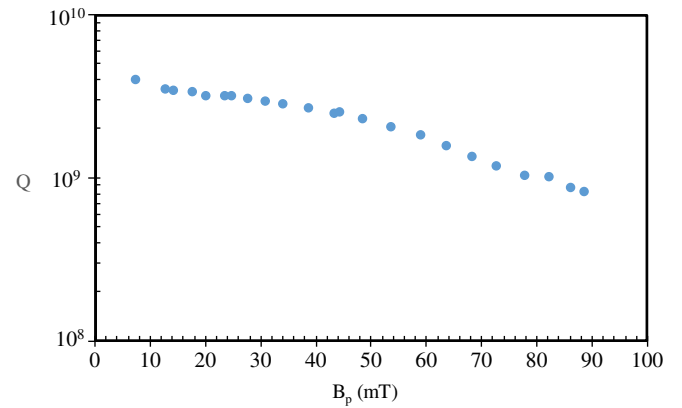

FIG. 21.   $Q$-curve for the 400 MHz rf-dipole cavity measured at 4.33 K.

Figure 21 shows a $Q$-curve measured at 4.33 K following the standard BCP chemical processing, a 600 C heat treatment for 10 hours, and a 120 C bake for 24 hours. From the geometrical factor $G = 107\,\Omega$ the average surface resistance $\bar{R}_s$ is obtained and shown as the dots in Fig. 22.

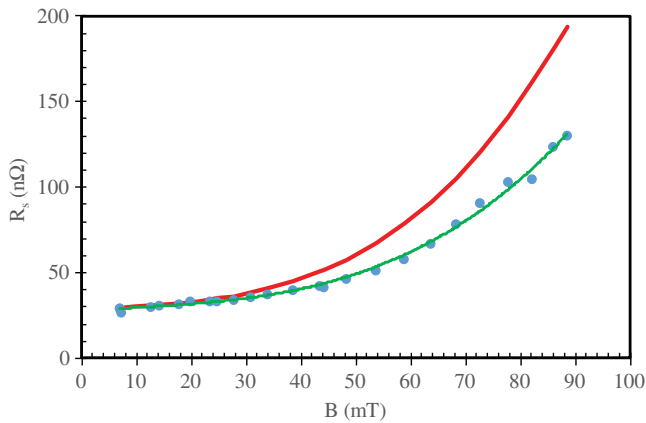

FIG. 22.  Experimental average surface resistance obtained from $Q$-curve in Fig. 16 (blue dots) with polynomial fit given by Eq. (27) (green line) and the derived real surface resistance given by Eq. (29) (red line).

A polynomial fit to the experimental data, shown as the green line going through the dots in Fig. 22, has the functional form

$$\bar{R}_s\left(\frac{B}{B_0}\right) = 27.8\left[1 + 0.77\left(\frac{B}{B_0}\right) \right.$$
$$\left. - 1.44\left(\frac{B}{B_0}\right)^2 + 6.00\left(\frac{B}{B_0}\right)^3\right], \quad (27)$$

where we have chosen $B_0 = 100$ mT and the surface resistance is expressed in n$\Omega$.

From Eq. (8) and Fig. 19 the correcting factors $\beta(\alpha)$ are

$$\beta(0) = 1, \quad \beta(1) = 1.24, \quad \beta(2) = 1.44, \quad \beta(3) = 1.63, \quad (28)$$

from which we obtain the actual surface resistance

$$R_s\left(\frac{B}{B_0}\right) = 27.8\left[1 + 0.95\left(\frac{B}{B_0}\right) \right.$$
$$\left. - 2.07\left(\frac{B}{B_0}\right)^2 + 9.76\left(\frac{B}{B_0}\right)^3\right]. \quad (29)$$

which is also shown in Fig. 22.

## VI. SUMMARY AND CONCLUSIONS

We have presented a general method to obtain to magnetic field dependence of a superconducting material from measurement of the $Q$-curve of a superconducting cavity. The method relies on a distribution function $a(h)$ (or its derivative) of the fraction of the cavity surface where the surface magnetic field is less than the fraction $h$ of the peak surface magnetic field. In a few cases $a(h)$ can be obtained analytically, more often it needs to be obtained numerically.

From the measurement of the "average" surface resistance, formulae have been presented relating the power expansion of the real surface resistance to that of the "average" surface resistance. The formulae are quite general in that the power expansions are arbitrary in size and not limited to integer powers. While the magnetic field dependence is obtained, the method still relies on the assumption that the superconductor has uniform properties over the whole surface.

The results have been applied to coaxial half-wave cavities, $TM_{010}$ cavities, and cavities of complex 3-D geometries. This method can also be straightforwardly applied to test cavities where the superconducting sample constitutes only a fraction of the whole system.

In the examples presented in this paper we have used polynomial expansions with integer exponents since they are relatively easy to obtain. As mentioned earlier our method is not restricted to integer exponents but can use any non-negative real numbers as exponents. For example we could have included half-integer exponents that could also have provided an excellent fit to the data but with a different power expansion. So, although this method can be expected to give reasonable numerical values for the actual surface resistance, caution should be exercised in drawing conclusions as to the functional dependence on the magnetic field. All that can be concluded is that the functional dependence is consistent with the assumed model.

## ACKNOWLEDGMENTS

This work was supported by National Science Foundation Grants No. PHY-1416051 and No. PHY-1632749. G. C. was supported at Jefferson Lab by U.S. DOE Contract No. DEAC05-06OR23177. This research used resources of the National Energy Research Scientific Center, which is supported by the Office of Science of the U.S. DOE under Contract No. DE-AC02-05CH11231.